\documentclass[twocolumn]{aa}
\usepackage{graphicx,natbib}
\usepackage{txfonts, multirow}
\begin{document}
   \title{Detection of the Sgr A* activity at 3.8 and 4.8 $\mu$m with NACO\thanks{Based on observations collected at the ESO VLT Yepun telescope, proposal 71.B-0365(A)}}

   \author{Y. Cl\'enet
          \inst{1}
          \and
          D. Rouan\inst{2}
          \and
          D. Gratadour\inst{2}
          \and
          F. Lacombe\inst{2}
          \and
          E. Gendron\inst{2}
          \and
          R. Genzel\inst{3}
          \and
          T. Ott\inst{3}
          \and
          R. Sch\"{o}del\inst{4}
          \and
          P. L\'ena\inst{2}
          }

   \offprints{Y. Cl\'enet}

   \institute{European Southern Observatory (ESO), Karl-Schwarzschild-Strasse 2, D-85748 Garching bei M\"unchen,
Germany\\
							\email{yclenet@eso.org}
         \and
   						Observatoire de Paris, LESIA, 5 place Jules Janssen, F-92195 Paris Cedex, France\\
              \email{\{daniel.rouan; damien.gratadour; francois.lacombe; eric.gendron; pierre.lena\}@obspm.fr}
         \and
              Max-Planck-Institut f\"ur extraterrestrische Physik, Giessenbachstrasse, D-85748 Garching bei M\"unchen, Germany\\
             \email{\{genzel; ott\}@mpe.mpg.de}
         \and
              I. Physikalisches Institut, Universit\"at zu K\"oln, Z\"ulpicher Strasse 77, D-50937, K\"oln, Germany\\
             \email{rainer@ph1.uni-koeln.de}
             }

   \date{Received ; accepted }

   \abstract{L'-band ($\lambda$=3.8~$\mu$m) and M'-band ($\lambda$=4.8~$\mu$m) observations of the Galactic Center region, performed in 2003 at VLT (ESO) with the adaptive optics imager NACO, have lead to the detection of an infrared counterpart of the radio source \object{Sgr A*} at both wavelengths. The measured fluxes confirm that the \object{Sgr A*} infrared spectrum is dominated by the synchrotron emission of nonthermal electrons. The infrared counterpart exhibits no significant short term variability but demonstrates flux variations on daily and yearly scales. The observed emission arises away from the position of the dynamical center of the \object{S2} orbit and would then not originate from the closest regions of the black hole.

   }

   \maketitle
%

\section{Introduction}
The conjugated increase in sensitivity and spatial resolution provided since the end of the 90's by adaptive optics (AO) systems on 8-10m class telescopes has highly raised the expectations to detect an infrared (IR) counterpart of \object{SgrA~*}, the radio source at the center of the Galaxy \citep{balick74}. The stellar proper motion studies performed for several years by two competing groups, first through speckle imaging then through AO at VLT and Keck, have made possible to trace orbits of several stars gravitationally bound to the central compact mass and to confirm the black hole nature of the latter \citep{schodel02,ghez03}.

Whatever mechanism is actually at work to produce the radio, submm and X  emission --  synchrotron emission from nonthermal electrons in a jet \citep{markoff01} or a shock, or synchrotron or even Bremstrahlung  from thermal electrons in the hot plasma of an ADAF disk  \citep{yuan02} --, an IR counterpart is predicted, at flux levels  which are within reach of current infrared imagers on large telescopes. Detecting this counterpart at several wavelengths is an important step to strongly constrain the  high frequency part of the spectrum and thus the details of the accretion mechanism on the black hole. Several attempts done during the past years in the L-band around 3.8 $\mu$m have remained unsuccessful or with uncertain results because of their lack of sensitivity \citep{forrest86,tollestrup89,depoy91,simons96,clenet01}. 

Despite a spatial resolution insufficient to separate the putative \object{Sgr~A*} IR emission from the one of \object{S2}, the closest star to the black hole, 2002 AO L'-band observations have lead to the detection of a possible emission from the black hole environment: from color excess derivation with NACO Science Verification data \citep{genzel03a,clenet04a}, from a comparison with \object{S2} 2003 photometric measurements with Keck AO data \citep{ghez04a}. In 2003, \object{S2} has sufficiently moved away and the very first direct detection of an IR emission coming from the black hole has been indeed observed in the H-, Ks- and L'-bands as short duration flares of 90 min typically \citep{genzel03b} and also as a more steady emission  \citep{clenet04b,ghez04a}. Together with studies at other wavelengths (eg, \citealp{bower04,eckart04}), both results brought a confirmation of a moderately active accretion process coupled to some mechanism, such as a jet, to produce nonthermal electrons. 

We report here the clear detection of an IR emission from the central black hole environment at both L' and M' (3.8 and 4.78~$\mu$m) which brings an additional constraint on the spectrum of the black hole emission.
  
\section{Observations and data reduction}
\subsection{Observations}
\label{section:obs}
Observations of the Galactic Center region have been performed with the 8m VLT UT4 Yepun telescope equipped with NACO, the NAOS adaptive optics system coupled to its IR camera CONICA \citep{lenzen98,rousset00}. L'-band (3.8 $\mu$m, 0.0271\arcsec/pixel) images have been obtained on 2 and 4 June 2003 and M'-band images (4.78 $\mu$m, 0.0271\arcsec/pixel) on 3 and 8 June 2003. Thanks to the NACO IR wavefront sensor and its ability to servo on \object{IRS~7}, the achieved spatial resolution, measured on \object{IRS~29N}, 4.4\arcsec\ away from the guide star, was close to the diffraction limit: 120, 132, 110, 132 mas for the June 2, 3, 4 and 8 images respectively.

At L', on-source and on-sky images have been alternatively acquired following an ABBA pattern, where the sky position was 3\arcmin away from the on-source one. A random jitter inside a 6\arcsec\ width box was applied on both on-source and on-sky positions. Each of the two individual 1024$\times$1024 pixel images obtained at each position resulted from the mean of 60 subintegrations of 0.175 s. An on-source image and its corresponding on-sky image were separated by about 75~s, for a total observing time of about 1.8~h on June 2 and 1.7~h on June 4.

At M', acquisitions have been performed combining secondary mirror chopping and telescope nodding. By chopping with a 11\arcsec\ or 15\arcsec\ throw to the North, an ABBA pattern has been followed to get on-source and on-sky images, the latter being then directly consecutive to the former. At each position, the resulting 512$\times$512 pixel image is the mean of 89 subintegrations of 0.056 s. Two successive ABBA patterns have been done before randomly jittering inside a 6\arcsec\ width box. The total observing time was about 3.2~h on June 3 and 2.8~h on June 8.

For both filters, the on-sky images have been subtracted from the corresponding on-source ones. The resulting images have been corrected from flat-field, then from bad pixels and finally from jittering by recentering them after a cross-correlation analysis. Data cubes of the Galactic Center region at different dates are thus obtained (after an image selection at L', based on an image quality estimation from the central flux of \object{IRS~29N}). For each observing night, the final L' and M' images have been built by applying a clipped mean on the time series of each pixel of the field. The resulting on-source integration times are 178.5 s (2 June) and 346.5 s (4 June) at L', and 224.3 s at M' (3 and 8 June). 

We also use L'-band data obtained on 30 August 2002, during NACO Science Verification (SV), which weren't available when writing our last article \citep{clenet04a} on the Galactic Center SV observations. Randomly jittered images have been collected within a jitter width box of 10\arcsec. The 1024$\times$1024 image recorded at each jittered position is the mean of 150 subintegrations of 0.2 s. The jittered images have been then corrected from flat-field and bad pixel. Each jittered image has been subtracted of its median value to account for large amplitude variations of the sky and stored as the successive planes of a data cube. A sky map has been built as follows: for each pixel, the sky value is computed by averaging the values along the third direction of the data cube after rejecting the lowest and the highest values in order to account for uncorrected bad and hot pixels and for stars. This sky map has been subtracted to each data cube image. These images have been then recentered. We have selected the images with the best image quality, estimated from the central flux of \object{IRS~29N} and finally averaged them. The final image corresponds to a total on-source integration time of 28 min.

\vspace{-1mm}

\subsection{Photometry}
\label{subsection:phot}
\begin{figure}[t]
   \begin{minipage}[c]{.49\linewidth}
      \includegraphics[width=5cm]{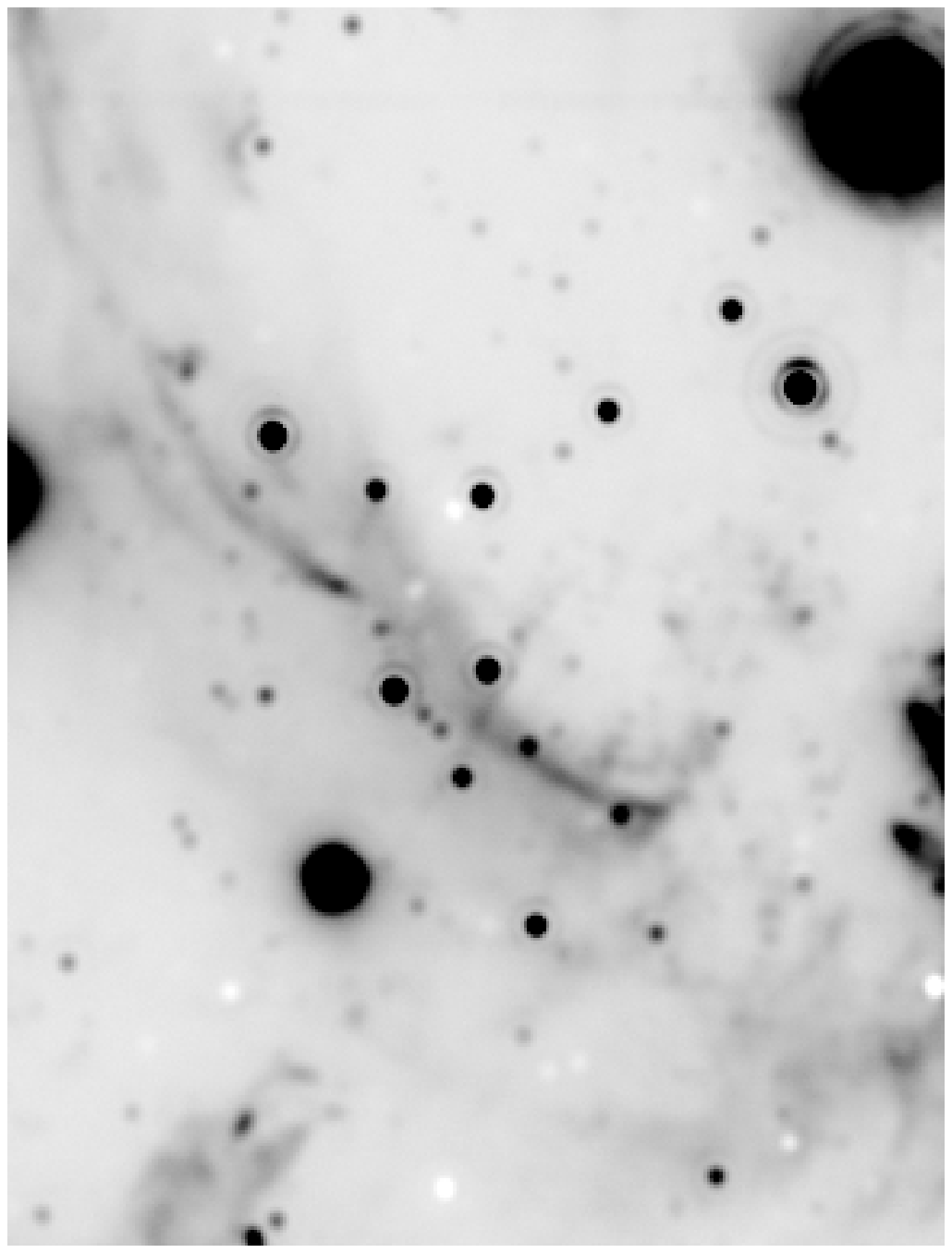}
   \end{minipage} \hfill
   \begin{minipage}[c]{.4\linewidth}
      \includegraphics[width=3.5cm]{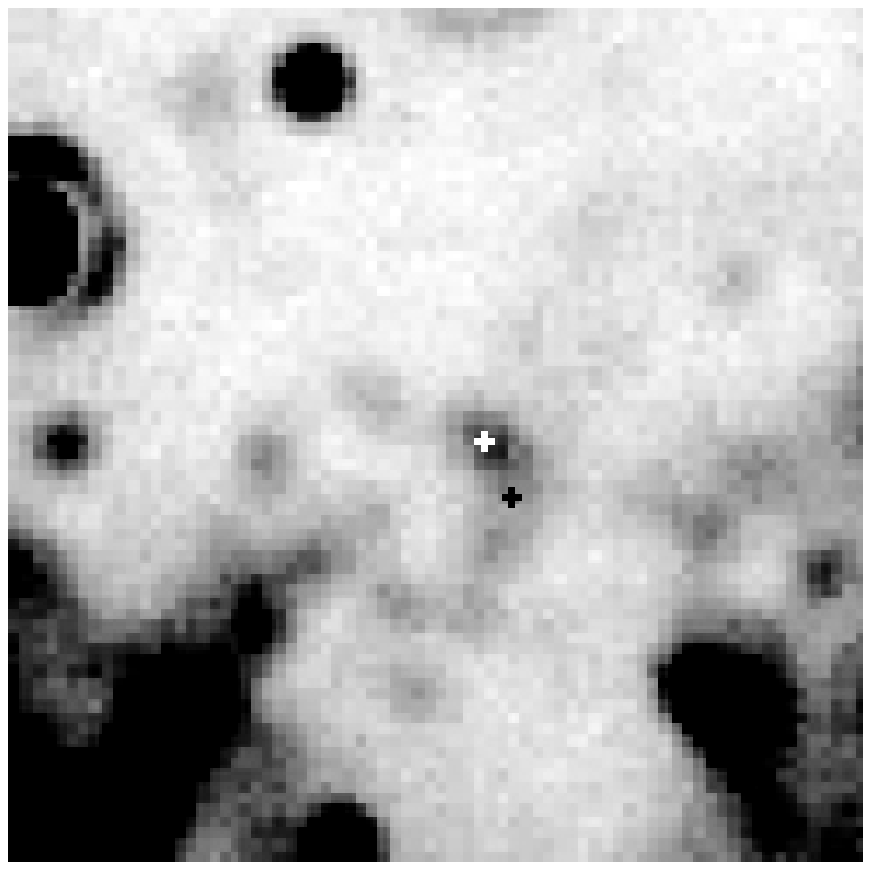}
   \end{minipage}
\caption{NACO images of the Galactic Center region at M' (2003 June 8). Left: the field of view is 7.9\arcsec$\times$10.5\arcsec. Right: A 2\arcsec$\times$2\arcsec\ close up on the Sgr~A* cluster. The white cross marks the position of \object{S2}, the black cross the position of Sgr~A*/IR.}
\label{fig:fig1}
\end{figure}

\begin{figure}[t]
   \begin{minipage}[c]{.49\linewidth}
      \includegraphics[width=5cm]{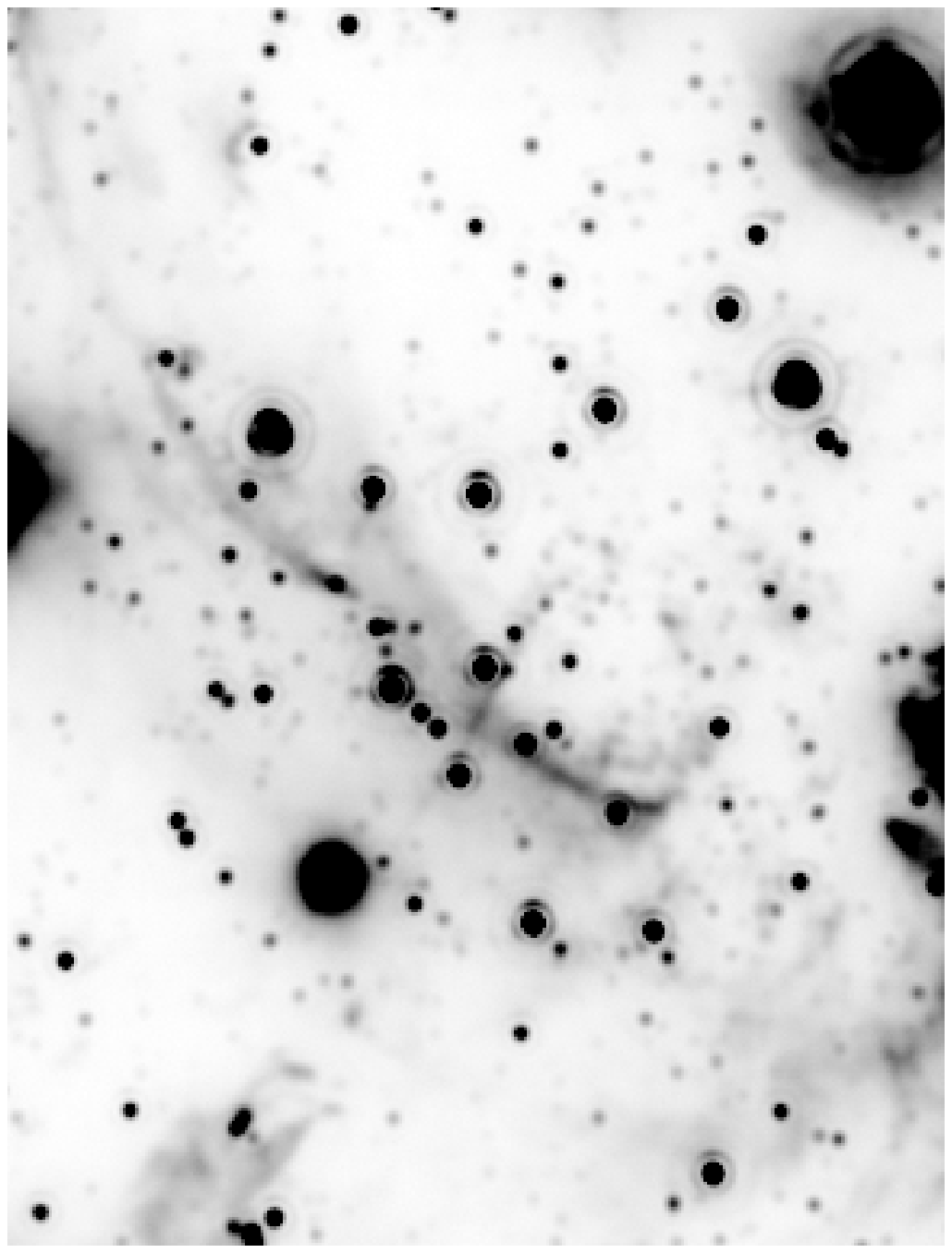}
   \end{minipage} \hfill
   \begin{minipage}[c]{.4\linewidth}
      \includegraphics[width=3.5cm]{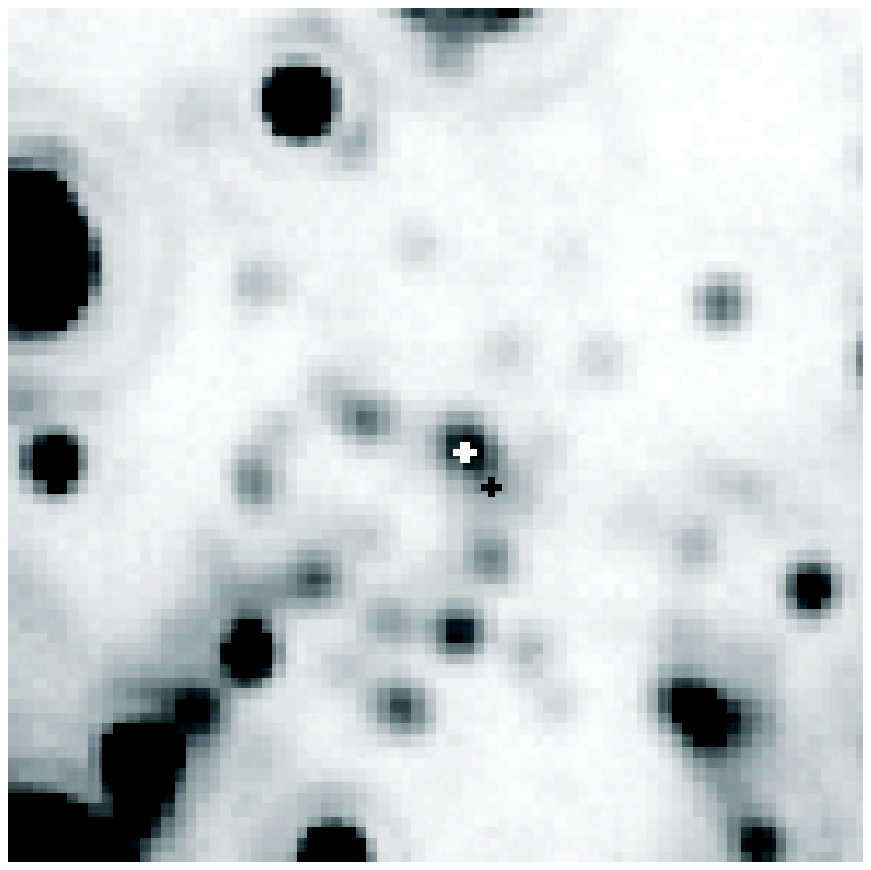}      

   \end{minipage}
\caption{Same as Fig.~\ref{fig:fig1} at L'(2003 June 4).}
\label{fig:fig2}
\end{figure}

The procedure followed to perform the absolute calibration from the relative L'-band photometry obtained with the PSF-fitting code Starfinder \citep{diolaiti00} has already been explained in \citet{clenet04a}. The non variable stars used for the calibration are \object{IRS~16C}, \object{IRS~29N} and \object{IRS~33SE}. Their absolute photometry is from \citet{blum96}. The M'-band calibration has been done assuming that the following four non variable (according to \citealp{ott99}) blue supergiants stars have a zero (L'$-$M')$_0$ index: \object{IRS~16NE}, \object{IRS~16NW}, \object{IRS~16C} and \object{IRS~33SE}. At L', we estimate the resulting photometric error to be 0.15 mag, 0.18 mag and 0.20 mag for the June 2, June 4 and the SV night respectively. At M', the estimated photometric error is 0.12 for the June 3 and 8 nights.

The L' zero-point offset found by \citet{ghez04a} between their photometry and the one of our previous work \citep{clenet04a}, L'$_\mathrm{Keck}$=L'$_\mathrm{Cl\acute{e}net}$+0.37, is persisting for the present work: from the mean \object{S2} 2003 photometry (L'=13.27 for \citealp{ghez04a} and L'=12.82 for our present work), we find now L'$_\mathrm{Keck}$=L'$_\mathrm{this\ work}$+0.45. This offset cannot come neither from the choice of the reference stars or from the absolute photometry of these reference stars: adopting the same reference stars (\object{IRS16 NE}, \object{IRS16 SW-E}, \object{IRS16 NW}, \object{IRS16 C}) and the same absolute photometry \citep{simons96} as \citet{ghez04a}, our L' zero point is offset by only $-$0.04 mag. A tentative explanation of this offset could be either (i) the difference in location of the sky positions; (ii) or the time delay between the acquisitions of the on-sky frames and the on-source ones: no precision is given on this point in \citet{ghez04a} and the time scale of sky emission variations can be significantly different from the one we had. Though, dereddened fluxes from the different works are directly comparable since this L' zero point offset is compensated by an equivalent difference of the extinction values: A$_\mathrm{L'}$=1.30 for this work (see below) and A$_\mathrm{L'}$=1.83 for \citet{ghez04a}.

By interpolating the extinction law values of \citet{moneti01}, which result for $\lambda$$>$2.5~$\mu$m from the modelisation of ISO SWS measurements \citep{lutz99}, and using A$_\mathrm{K}$=2.7 \citep{clenet01}, we obtain A$_\mathrm{L'}$=1.30 and A$_\mathrm{M'}$=1.21. Dereddened fluxes are computed assuming zero magnitutes values from \citet{cox00}: F$_0$(L')=248~Jy and F$_0$(M')=160~Jy. A distance to the Galactic Center of 7.94 kpc was assumed \citep{eisenhauer03}.

\begin{table}[t]
\centering \caption[]{Astrometry and photometry of Sgr~A*/IR and \object{S2}. Offsets are given in mas relatively to the dynamical center of the \object{S2} orbit. Values in bracketts are the offsets expressed in units of the diffraction limit. 2002 photometry has been computed assuming for \object{S2} the 2003 weighted mean photometry. Ref -- Gh: \citet{ghez04a}, Cl1: \citet{clenet04a}, Cl2: this work, Ge: \citet{genzel03b}. Sgr~A*/IR and \object{S2} being superimposed in 2002, the offsets have been set to 0.}
\label{table:position}
\resizebox{\hsize}{!}{\begin{tabular}{l c c c c c c}
\hline
\hline
\multirow{2}{0.3cm}{Date} & \multirow{2}{0.8cm}{Filter} & \multicolumn{3}{c}{Sgr A*/IR}  & \object{S2} & \multirow{2}{0.5cm}{Ref}\\
 & & $\Delta\alpha$ & $\Delta\delta$ & mJy & mag & \\
\hline
2002 May  31 & L' & 0$\pm$6 & -7$\pm$7 & 8.2 $\pm$0.6 & 13.27 & Gh\\
2002 Aug. 19 & L' & 0$\pm$3 & 0$\pm$3 & 14.7$\pm$6.7 & 12.82$\pm$0.12 & Cl1\\
2002 Aug. 30 & L' & 0$\pm$3 & 0$\pm$3 & 14.3$\pm$3.8 & 12.82$\pm$0.12 & Cl2\\
2002 Aug. 30 & L' & 0$\pm$30 & 0$\pm$30 & 17.5$\pm$5 & 12.92 & Ge\\
2002 Aug. 30 & L' & 0$\pm$30 & 0$\pm$30 & 30.1$\pm$4 & 12.92 & Ge\\
2003 May   9 & L' & -9$\pm$15 & -4$\pm$20 & 6.4$\pm$1.9 & 12.92 & Ge\\
2003 Jun. 2 & L' & $-$12$\pm$21 (0.12) & $-$11$\pm$23 (0.11) &  3.2$\pm$0.5 & 12.80$\pm$0.15 & Cl2\\
2003 Jun. 3 & M' & $-$51$\pm$19 (0.41) & $-$5$\pm$23 (0.04) &  4.5$\pm$0.5 & 12.38$\pm$0.12& Cl2\\
2003 Jun. 4 & L' & $-$32$\pm$36 (0.33) & $-$10$\pm$17 (0.10) & 3.1$\pm$0.4 & 12.86$\pm$0.18& Cl2\\
2003 Jun. 8&  M' & $-$27$\pm$25 (0.22) & $-$43$\pm$42 (0.35) &  3.5$\pm$0.4 & 12.36$\pm$0.12 & Cl2\\
2003 Jun. 10 & L' & -8$\pm$9 & -1$\pm$10 & 16.4$\pm$0.8 & 13.27& Gh\\
2003 Jun. 16 & L' & -15$\pm$9 & 9$\pm$8 & 12.9$\pm$0.6 & 13.27& Gh\\
2003 Jun. 17 & L' & -12$\pm$13 & -7$\pm$18 & 5.9$\pm$0.2 & 13.27 & Gh\\
\hline
\end{tabular}}
\label{table:phot}
\vspace{-0.5mm}
\end{table}

\vspace{-1mm}

\subsection{Astrometry}
The astrometry, also obtained from Starfinder, has been performed relatively to the dynamical center of the \object{S2} orbit: using the \object{S2} orbital parameters \citep{eisenhauer03}, we have computed the offsets between \object{S2} and the dynamical center (eg, $\Delta\alpha$=36.1~mas, $\Delta\delta$=75.5~mas for 2003 June 4) to localize the latter on our images. Adopting the parameters of \citet{ghez04b} would have shifted the astrometry of only $-2$~mas in right ascension and $+3$~mas in declination. Assuming gaussian distributions for the \object{S2} orbital parameters, simulating several \object{S2} orbits leads to uncertainties of 3 mas in right ascension and declination for the offset between \object{S2} and the dynamical center.

Astrometric errors in Table~\ref{table:phot} result from the uncertainties of the offset between \object{S2} and Sgr~A*/IR, computed by running Starfinder on subdivided data cubes, and from the uncertainties of the offset between \object{S2} and the dynamical center (see above).

\vspace{-1mm}

\section{The L'- and M'-band emission from \object{Sgr A*}}
\subsection{Detection of Sgr A*/IR}
In 2002, the angular resolution delivered by AO systems on 8-10m telescopes was not sufficient to spatially separate \object{Sgr~A*} from \object{S2}, the star at closest approach. In 2003, despite a \object{Sgr~A*}-\object{S2} distance (85~mas) still smaller than the achieved spatial resolution (cf. Sect.~\ref{section:obs}), this was no longer the case: on {\it all} our L' and M' images, Starfinder detects south west to \object{S2} a second source with a L' dereddened flux of 3.2$\pm$0.5~mJy on June 2, 3.1$\pm$0.4~mJy on June 4, and a M' dereddened flux of 4.5$\pm$0.5~mJy on June 3, 3.5$\pm$0.4~mJy on June 8 (Table~\ref{table:position}). This additional source appears to be located south-west from the dynamical center. The longer the wavelength, the larger the distance: 16$\pm$22 mas and 34$\pm$35 mas at L', 52$\pm$19 mas and 51$\pm$38 mas at M'. This is not unexpected because the confusion with \object{S2} is less severe at M' relative to L' thanks to the better colour contrast. Then the Sgr~A*/IR astrometry is less affected by the \object{S2} brightness at M' compared to L'.

For the following reasons, we claim that this
source is most probably the IR counterpart of \object{Sgr~A*}, whose detection has been also reported from IR observations in 2003 \citep{genzel03b,clenet04b,ghez04a}:
\begin{itemize}
\item the dispersion of the positions is small: for instance, the largest distance between measured locations of this second source is typically one third of the corresponding diffraction limit; 
\item this second source is very red with an intrinsic  L'$-$M' color index larger than 0.65, a value much too high to be explained by a background star;
\item this source demonstrates a significant variability on yearly scale at L' (between 2002 and 2003 photometry) and even on day scale at M' (between the 2003 June 3 and 8). Note that Sgr~A*/IR 2002 photometry has been computed by subtracting the contribution of \object{S2}, assuming that it is non-variable and correctly measured with the 2003 data.
\end{itemize}

\vspace{-1mm}

\subsection{Spectrum of  Sgr A*/IR}
\label{section:spectrum}

\begin{figure*}[t]
\begin{center}
\includegraphics[scale=0.52]{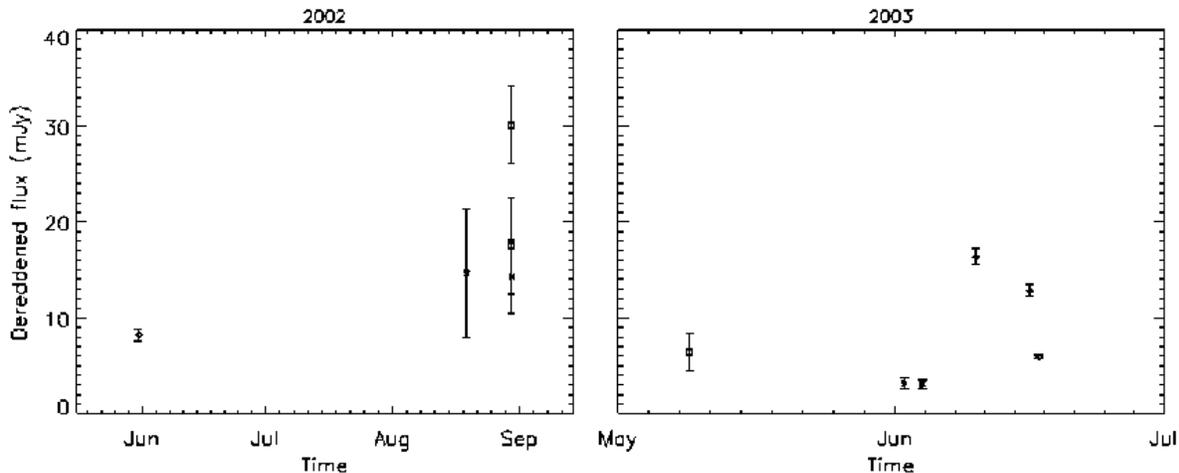}
\end{center}
\vspace{-3mm}
\caption{L'-band dereddened flux of Sgr~A*/IR since mid 2002, from the gathering of all published measurements: this work (stars), \citet{genzel03b} (squares) and \citet{ghez04a} (diamonds).}
\label{fig:variable}

\end{figure*}

The clear detection of the IR emission of \object{Sgr~A*} at L'  (\citealp{genzel03b,clenet04a,clenet04b,ghez04a} and this work) must be considered as an important first step: the flux level agrees reasonably well with the prediction for a synchrotron emission from hot electrons  \citep{yuan03,yuan04} and is far above the Bremsstrahlung emission of a standard ADAF model. This result tends then to support the claim that the same population of nonthermal electrons is responsible for both the excess radio emission at low frequency ($\nu$$<$10$^{10.5}$~Hz) and the IR emission, while the sub-mm bump and the steady state X ray emission would come respectively from synchrotron emission by thermally distributed electrons and its inverse Compton component.

The exact origin of the nonthermal electrons component is not totally certain and models either point at the nozzle of a jet where a shock delivers the energy \citep{markoff01} or at some MHD turbulence phenomenon \citep{yuan03,yuan04}, but in all cases, only few per cent of the energy should be in this power law tail of nonthermal electrons.
   
The constraint brought by our new measurements is thus  of great importance to assess the nonthermal electrons model. If we first consider the quiescent emission model of \citet{yuan03}, the agreement appears fairly good, especially at L': $\nu$\,L$_\nu$=1.90$\times$10$^{34}$~erg~s$^{-1}$, below the theoretical curve by a factor 1.2. At M', the average luminosity ($\nu$\,L$_\nu$=1.90$\times$10$^{34}$~erg~s$^{-1}$) is below the theoretical curve by a factor of 1.5.

 Our measurements indicate a flat spectrum in $\nu$\,L$_\nu$, while \citet{yuan03} predict a slope of synchrotron emission of $\sim$0.77 around 4 $\mu$m. If we assume no significant variation during our observations, we can explore the compatibility of a flat spectrum with the models. The emission from thermal electrons is excluded since the predicted flux is much below the observed one and the slope is in addition too steep (Fig.~5 in \citealp{yuan03}). A flat spectrum at 4~$\mu$m is predicted for a power-law index of the nonthermal electrons distribution of 2.5 (Fig.~4 in \citealp{yuan03}), but the predicted flux is too high at 4~$\mu$m, by a factor $\sim$5, and in the X-rays. The model and the observations would agree if only 0.3~$\%$ of the energy were in the nonthermal component, instead of 1.5~$\%$, but the low frequency radio part of the spectrum could then not be fitted.

More recently, \citet{yuan04} have adjusted the parameters of their model to mainly account for the H- and Ks-band quiescent emission of \object{Sgr~A*}. Though, this updated model underestimates the L'-band fluxes considered in their work, which are among the highest values from \citet{genzel03b} and \citet{ghez04a}, and overestimates our quescient L' and M' measurements more than their precedent model.

 A significant variation of the emission between the observations at L' and M' would more easily explain the difference in slope between our observations the modelled spectrum: an even larger magnitude of variation has already been observed in a single night by \citet{ghez04a}. In the future, obtaining simultaneous IR measurements will be extremely valuable to assess both the variability behaviour and to constrain the nonthermal electron population.

\vspace{-2mm}

\subsection{Variability of Sgr A*/IR}

During a night, our 2003 L' and M' measurements exhibit no significant short time scale variability or periodicity, on the contrary to what has been observed at L' by \citet{genzel03b} and \citet{ghez04a}. Though, collecting all the L'-band flux measurements of Sgr~A* published so far (Table~\ref{table:phot}) demonstrates that the black hole environment can experience three types of variability in this wavelength range (Fig.~\ref{fig:variable}): (i) on a short time scale, typically 30 min (2002 Aug. 30 flare, \citealp{genzel03b}), similarly to the near-IR and X-rays observations, with a flux amplification of a factor $\sim$1.5, (ii) on a day time scale, as shown by the burst observed between 2003 June 2 and 17 (\citealp{ghez04a} and this work) where the flux varied by a factor $\sim$5, (iii) and on a year time scale with an amplification factor from 2.5 to 4.5 as observed between the 2002 and 2003 quiescent fluxes. In addition, we have observed a significant variation in the M' photometry of Sgr~A*/IR with a reduction of more than 20\% of its flux in 5 days.

At L', the shortest time scale variability appears then to have the lowest amplitude and should be related to the synchrotron emission of non thermal electrons accelerated in the first few Schwarzschild radii of \object{Sgr~A*}, as confirmed by the good agreement between our flux measurements and the corresponding models (Sect.~\ref{section:spectrum}).

 Concerning the longer time scale variability, a star close to the black hole, passing through an inactive accretion disk, as proposed by \citet{nayakshin03} to explain the X-ray flares, would hardly account for the bright burst observed in June 2003: the authors claim that the typical duration of a flare would be a few tens of kiloseconds, much shorter than the observed burst, and that the longer the flares the weaker they are. 

Similarly, a variability of \object{S2} could be responsible for the yearly variability of Sgr~A*/IR we have observed between 2002 and 2003: to compute the 2002 L' photometry of Sgr~A*/IR, we have assumed that the \object{S2} contribution to the unresolved \object{S2}+Sgr~A*/IR source was its averaged 2003 magnitude. \citet{cuadra03} have shown that the eclipse of \object{S2} on its orbit by an inactive accretion disk could result in \object{S2} flux variations at L' but not at H or Ks (as reported in the literature). Though, this L' flux variation would not occur at the observed date, before mid 2001 instead of mid 2002, and the inner radius of this solution ($R_{in}$=0) would differ from the value adopted to explain the \object{Sgr~A*} X-ray flares \citep{nayakshin03}. An exploration of the parameter space may account for all these observational constraints.

The longer time scale variability should then be intrinsic to Sgr~A*/IR. The dissimilar characteristics (intensity, time scale) of this variability compared to the shortest one suggest it could be related to a mechanism different from the one invoked for the short flares: either a long term enhanced accretion or a variability in the jet emission through some injection of electrons.
 
\vspace{-1mm}
 
\subsection{Position of Sgr A*/IR}
Till now, it has been claimed that all IR detections of Sgr~A* were directly related to the closest parts to the black hole. Though, for our four observing nights, whatever the filter is, the second source detected by Starfinder is offset to the west and more negligibly to the south with respect to the dynamical center of the \object{S2} orbit (Table~\ref{table:position}). The previous detections of Sgr~A*/IR \citep{genzel03b,ghez04a} show a similar trend but with lower offset values (Table~\ref{table:position}). 

This offset is of the same order as the corresponding astrometric error at L' (16$\pm$22 mas and 34$\pm$35 mas) and a bit larger at M' (52$\pm$19 mas and 51$\pm$38 mas). Its measurement has therefore a rather small degree of confidence (down to 53\% at L', 82\% at M') and may not be significant. It could result from the still closeliness of \object{Sgr~A*} to \object{S2} in 2003 (84.6~mas, 86\% of the diffraction limit at L', 69\% at M') conjugated to the low contrast between the two sources. 

Though, if confirmed, this overall offset would trace an emission far away from the inner parts of the black hole: the mean location at L' is at 3$\times$10$^3$ Schwarszchild radius ($R_\mathrm{S}$) of the black hole, 6$\times$10$^3$~R$_\mathrm{S}$ at M'. These emissions could still come from the accretion disk since its outer radius (the Bondi accretion radius) is about 10$^5$~R$_\mathrm{S}$ but the large variation of flux observed at L' during the 2003 June burst may hardly originate from a region with such a weak density. Alternatively, the emission could come from the interaction of a jet with the material surrounding the black hole and the difference of emission locations between L' and M' could then trace the different temperatures of the gas and dust heated by the jet. The shift of \object{S2} on its orbit around \object{Sgr~A*} should give the opportunity to assess  this putative offset in 2004.

\end{document}